\documentclass[cits]{PoS}

\usepackage{graphicx}
\usepackage{amsmath,amssymb,enumerate,epsfig,subfigure,mathrsfs}

\DeclareMathAlphabet{\mathpzc}{OT1}{pzc}{m}{it}

\def\msC{{\mathscr C}}

\def\calY{{\mathcal Y}}
\def\funY{{\mathpzc Y}}

\def\traza{{\rm Tr}}

\def\a{\alpha}

\def\ga{\gamma}

\def\la{\lambda}

\def\m{\mu}
\def\n{\nu}

\def\s{\sigma}

\def\th{\theta}
\def\eps{\epsilon}
\def\ee{\varepsilon}

\def\aslash{{a\mkern-9mu/}}

\def\Dirac{{D\mkern-12mu/}}
\def\Dp{{\mathcal{D}\mkern-12mu/}\,}

\def\prslash{{\partial\mkern-9mu/}}

\def\prslash{{\partial\mkern-9mu/}}    %%_standard_Dirac_operator

\def\muo{\mu_1}
\def\mutw{\mu_2}
\def\muth{\mu_3}
\def\mufo{\mu_4}

\def\id{{\rm{I}\!\rm{I}}}

\def\tpsi{\widetilde \psi}
\def\tphi{\widetilde \phi}
\def\tPsi{\widetilde \Psi}
\def\tPhi{\widetilde \Phi}
\def\tlambda{\widetilde \lambda}
\def\tLambda{\widetilde \Lambda}
\def\tSigma{\widetilde \Sigma}
\def\idp{\int\!\! \frac{d^4\!p}{(2\pi)^4}}

\def\idq{\int\!\! \frac{d^4\!q}{(2\pi)^4} \,\,}

\def\id3x{\int\!\! d^3\!\vec{x}}
\def\idx{\int\!\! d^4\!x}

\def\rig>{\right>}
\newcommand{\bea}{\begin{eqnarray}}
\newcommand{\eea}{\end{eqnarray}}
\newcommand{\beann}{\begin{eqnarray*}}
\newcommand{\eeann}{\end{eqnarray*}}
\newcommand{\ba}{\begin{array}}
\newcommand{\ea}{\end{array}}

\newcommand{\f}[3]{{f_{#1#2}}^{#3}}

\newcommand{\Tr}{\mathbf{Tr}}

\def\Psib{\bar{\Psi}}
\def\g5{\gamma_{5}}

\def\Dp{{\mathcal{D}\mkern-12mu/}\,}
\def\Rs{{R\mkern-11mu/}\,}
\def\prslash {{\partial\mkern-9mu/}}  %operador Dirac

\def\idx3{\int\! d^{3}\!\vec{x}\,}
\def\idx{\int\! d^{4}\!x\,}

 \def\Psib{\bar{\Psi}}

 \def\aslash{{a\mkern-9mu/}}
 \def\Dirac{{D\mkern-12mu/}\,}
 \def\hatDirac{{\hat{D}\mkern-12mu/}\,}
 \def\Dp{{\mathcal{D}\mkern-12mu/}\,}
 \def\Rs{{R\mkern-11mu/}\,}
 
 \def\prslash {{\partial\mkern-9mu/}}  %operador Dirac

 \def\g {\gamma}
 \def\a {\alpha}

 \def\s {\sigma}

\def\eps{\epsilon}
\def\epsb{\bar{\eps}}
\def\la{\lambda}
 \def\Tr{\text{Tr}}

\def\limit{\lim_{\Lambda\rightarrow\infty}}
\def\id{{\rm{I}\!\rm{I}}}
\def\m{\mu}
\def\n{\nu}
\def\s{\sigma}
\def\sb{\bar{\sigma}}
\def\g{\gamma}
\def\ga{\gamma}

\title{NC GUTS: A Status Report}

\ShortTitle{NC GUTS: A Status Report}

\author{\speaker{C.P. Martin}\\
         Departamento de F\'{\i}sica Te\'orica I\\
        Universidad Complutense de Madrid\\
         Madrid-28040, Spain\\
E-mail: \email{carmelo@elbereth.fis.ucm.es}}

\abstract{I review the main results that have been obtained so far on
the construction of noncommutative GUTs}

\FullConference{Corfu Summer Institute on Elementary Particles and Physics - Workshop on Non Commutative Field Theory and Gravity\\
                 September 8  12, 2010\\
                 Corfu, Greece}

\begin{document}

\section{Introduction}
 It is already ten years since the publishing of refs.~\cite{Madore:2000en,
   Jurco:2000fs, Jurco:2000ja}, where it was put forward a formalism --called
the enveloping-algebra formalism-- which led to refs.~\cite{Calmet:2001na,
  Aschieri:2002mc}, where the Noncommutative Standard Model and noncommutative
GUTs were formulated. An excellent introduction to
noncommutative gauge theories defined within the enveloping-algebra formalism
can be found in ref.~\cite{ Blaschke:2010kw}.

Let us recall that in the enveloping-algebra formalism the noncommutative
fields are functions of the ordinary fields --ie, no change in the number of
degrees of freedom as we move from ordinary to noncommutative space-time--
such that ordinary gauge orbits are mapped into noncommutative gauge orbits:

\begin{equation}
\begin{array}{l}
{A_\m[a_\m,\psi,\theta]+s_{NC}\,A_\m[a_\m,\psi,\theta]A=A_\m[a_\m+s\,a_\m,\psi+s\,\psi,\theta]},\\
{\Psi[a_\m,\psi,\theta]+s_{NC}\,\Psi[a_\m,\psi,\theta]=\Psi[a_\m+s\,a_\m,\psi+
s\,\psi,\theta],}\\
{s_{NC}\Lambda[\lambda,\lambda,\psi,\theta]=s\Lambda[\lambda,\lambda,\psi,\theta]},\\
[4pt]
{A_\m[a_\m,\psi,\theta=0]=a_\m,
\Psi[a_\m,\psi,\theta=0]=\psi,\Lambda[\lambda,\lambda,\psi,\theta=0]=}\lambda
\\[4pt]
{s_{NC}\,A_\m=\partial_\m\Lambda-i[A_\m,\Lambda]_{\star},
s_{NC}\,\Psi=i\Lambda\star\Psi,
s_{NC}\,\Lambda=i\Lambda\star\Lambda},\\
{s\,a_\m=\partial_\m\lambda-i[a_\m,\lambda],s\,\psi=i\lambda\,\psi,
s\,\lambda=i\lambda\,\lambda},
\end{array}
\label{standard}
\end{equation}

I shall call these equations standard Seiberg-Witten map equations since
$\Lambda$ acts from the left on the matter fields $\Psi$. The solution
to these equations which match the corresponding ordinary field when the
noncommutativity matrix, $\theta^{\mu\nu}$, vanishes shall be called standard Seiberg-Witten map. Now,
since   $a_\m$ and $\lambda$ take values on the Lie algebra, $\mathfrak{g}$, of a
compact Lie group, G, then, the noncommutative fields $A_\m$ and $\Lambda$ take
values on the universal enveloping algebra of $\mathfrak{g}$. This is a
characteristic feature of noncommutative gauge fields defined in the
enveloping-algebra formalism.

Having defined the noncommutative gauge and matter fields in terms of the
ordinary fields using the solution to eq.~(\ref{standard}), we now introduce de classical
(nonSUSY) noncommutative GUT(-inspired) theory for a compact Lie group, G,
by giving its action $S$:

\begin{equation}
\begin{array}{l}
{S=S_{gauge}\,+\,S_{fermionic}\,+\,S_{Higgs}\,+\,S_{Yukawa}},\\[4pt]
{S_{gauge}\idx-\frac{1}{2}\sum_{{\cal R}}\,c_{{\cal R}}\Tr_{{\cal R}} F_{\m\n}[{\cal R}(A)]\star F^{\m\n}[{\cal R}(A)]},\\[4pt]
{S_{fermionic}=\idx\Psib_L i\Dirac[\rho_{\psi}(A)] \Psi_L},\\[4pt]
S_{Higgs}\quad\text{and}\quad S_{Yukawa}\quad\text{give to},\\[4pt]
{F_{\m\n}[{\cal R}(A)]=\partial_\m {\cal R}(A)_\n-\partial_\n {\cal R}(A)_\m-i[{\cal R}(A)_\m, {\cal R}(A)_\n]_\star},\\
{D_{\m}[\rho_{\psi}(A)]\psi_L=\partial_\m\Psi_L-i
\rho_{\psi}(A_\m)\star\Psi_L}.
\end{array}
\label{classaction}
\end{equation}

$S_{Higgs}$ and $S_{Yukawa}$ yield, respectively, the Higgs and Yukawa
sectors of the GUT theory and are dropped to define what we call
noncommutative GUT-inspired theories. We shall see later on that the
construction of $S_{Yukawa}$ is far from trivial and it demands the use
of the so-called Hybrid Seiberg-Witten maps~\cite{Schupp:2001we} --needed to define noncommutative
gauge transformation acting from the left and from right. In
eq.~(\ref{classaction}), $\Psi_L[\theta^{\mu\nu},\rho_{\psi}(a),\psi_L]$ is
the noncommutative left-handed spinor multiplet which is the noncommutative counterpart
of the ordinary left-handed spinor multiplet $\psi_L$. $\psi_L$ carries an
arbitrary unitary representation, $\rho_{\psi}$, of $\mathfrak{g}$. ${\cal R}$
labels the unitary irreps --typically the adjoint and matter irreps-- of
$\mathfrak{g}$ and $\sum_{\cal R}c_{\cal R}\Tr_{\cal R}{\cal R}(T^a_{\text{I}}){\cal R}(T^a_{\text{I}})=1/g_{\text{I}}^2$.

Next, the quantum version of the classical field theory defined above is
obtained by integrating over the ordinary fields in the path-integral with
Boltzmann factor $e^{i\,S}$. $S$ is the action above, which we shall
understand as a formal power series in $\theta^{\mu\nu}$. I believe that this
expansion in $\theta$ will not yield the right Physics at
$\mbox{Energies}\,>\, 1/\sqrt{\theta}$.

After those ten years, it is advisable that we pause to look back and
assess what has been achieved as regards the quantum properties of those
GUT(-inspired) theories. I will not cover all that has been done so far, but
I will focus on

\begin{itemize}

\item Gauge anomalies.
\item Renormalisability (when there are no Higgs and no Yukawa sectors),
\item Construction of Yukawa terms.
\item Existence of Supersymmetric versions.

\end{itemize}

\section{Gauge Anomalies}

When quantising a chiral gauge theory the first problem one has to face is
that of gauge anomalies, for if the latter exist the theory will not make sense at
the quantum level. The chiral vertices in the classical action acquire
$\theta$-dependent terms, which can give rise to new $\theta$-dependent
anomalous contributions to the famous ordinary triangle diagrams:

\begin{equation*}
S_{fermionic}=\idx\,\bar{\psi}i\prslash\psi
+\bar{\psi}\{\aslash\,-\theta^{\alpha\beta}
[\frac{1}{2}f_{\alpha\beta} i\Dirac\,(a)+
\gamma^{\rho}f_{\rho\alpha}iD_{\beta}\,(a)]\}\,{\rm P}_{L}
\psi\,+\, o(\theta^2).
\end{equation*}

Thus, I started long ago the computation of the  three types of one-loop
three-point diagrams in Figure 1 giving would-be $\theta$-dependent anomalies.

\begin{center}
\epsfig{file=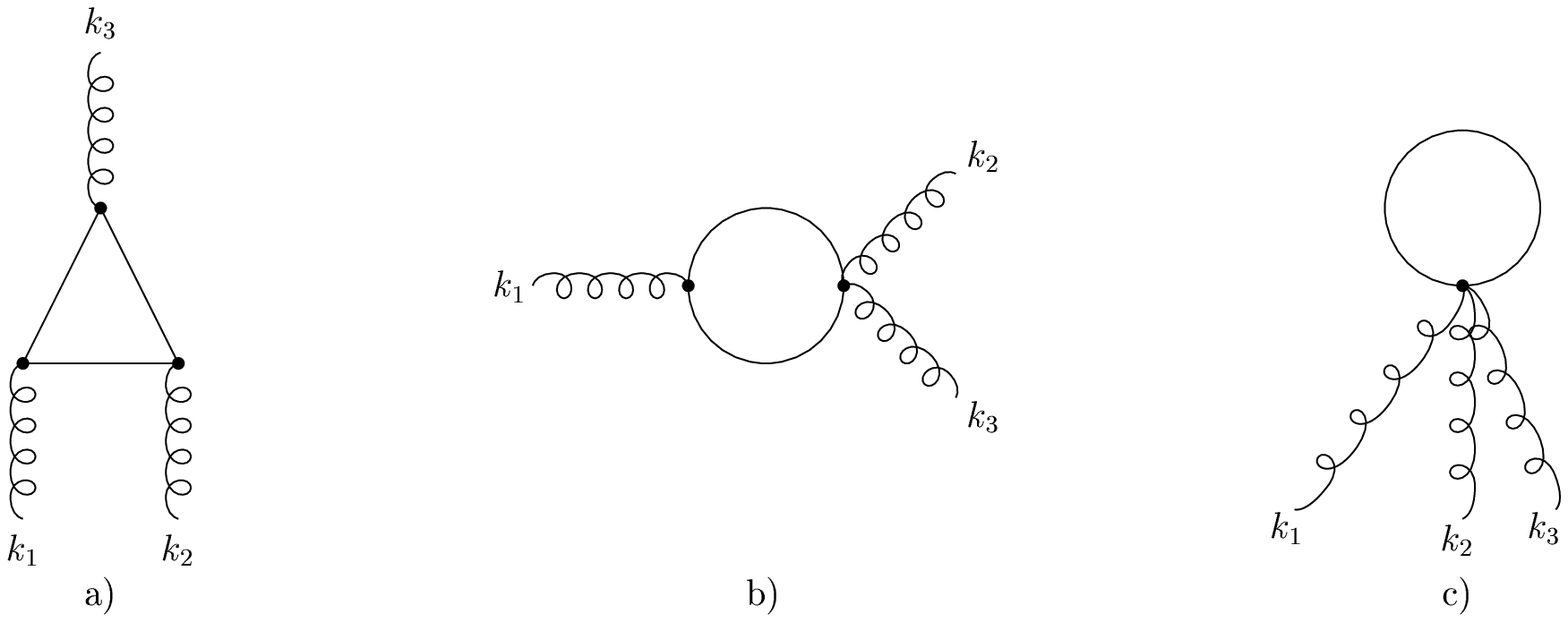,width=.60\textwidth}\\[20pt]
{ {\bf Figure 1}: Types of $\theta$-dependent would-be anomalous
three-point diagrams.}
\end{center}
\noindent
Actually, I was completely sure that these diagrams would give rise to new
$\theta$-dependent anomalous terms, which would lead to extra anomaly
cancelation conditions, which in turn would make most --eg., the Noncommutative Standard Model,
noncommutative GUTs,...-- of these theories meaningless at the quantum level. I could not
be more mistaken! I was very surprised to find that the $\theta$-dependent
anomalous contributions to the 1PI functional, $\Gamma$, were BRS-exact. ie,
they were not truly anomalous terms:
\begin{equation}
\begin{array}{l}
{s\Gamma[A[a,\theta],\theta]=-\frac{i}{24\pi^2}\idx\,
\ee^{\muo\mutw\muth\mufo}\,\traza\big(\partial_{\muo}\lambda\,
a_{\mutw}\partial_{\muth}a_{\mufo}\big)}\\[4pt]
+
 {s\Big[\frac{1}{48\pi^2}\,\idx\,
\ee^{\muo\mutw\muth\mufo}\,\theta^{\alpha\beta}\,\traza\,\big(
\partial_{\alpha}\partial_{\muo}
a_{\mutw}\partial_{\muth} a_{\mufo}a_{\beta}\big)\,\Big]+o(a^3)+o(\theta^2).}
\end{array}
\label{ganom}
\end{equation}

The computations that led to the previous results were carried out by using
dimensional regularization with a nonanticommuting $\gamma_5$. More details
can be found un ref.~\cite{Martin:2002nr}. I would like to point out now that
when I did the computations back in 2002, I was completely unaware of the
results --obtained using cohomological techniques-- by Barnich, Henneaux and
Brandt on the lack on nonBardeen anomalies for semisimple Lie algebras. The
result in eq.~(\ref{ganom}) holds, though, for nonsemisimple Lie algebras as well.

The next challenge was to show, at one-lop, that there were no
$\theta$-dependent gauge anomalies at any order in $\theta$ and for any number
of $a_\m$'s. We did so --see ref.~\cite{Brandt:2003fx}, for details-- by using
a mixture of explicit Dimensional Regularization computations, brute force of
BRS equations and cohomological BRS techniques. Indeed, by taking  advantage of the fact
that in Dimensional Regularization the Jacobian of ${\cal J}=\mathbb{I}+M$ --an
operator which enters the Seiberg-Witten map for fermions
   $ \Psi_{\alpha I} = \big( \delta_{IJ}\,\delta_{\alpha\beta} +
     M[a,\partial,\gamma,\gamma_5;\theta]_{\alpha\beta\;IJ}\big)\, \psi_{\beta J}$--
 is trivial, we were able to obtain the complete gauge anomaly candidate:
\begin{equation*}
\begin{array}{l}
 {{\cal A}[A,\Lambda,\theta] = - \frac{i}{24\pi^2}
         \idx\epsilon^{\m_1\m_2\m_3\m_4}\mathbf{Tr}\>
              \Lambda\! \star\! \partial_{\m_1}\Big(
                 A_{\m_2}\star\partial_{\m_3} A_{\mu_4}
       + \frac{1}{2} A_{\m_2}\!\star\! A_{\m_3}\!\star\! A_{\m_4}\Big),}\\[4pt]
{A_\m =A[a,\theta]_\m,\quad\Lambda=\Lambda[\lambda,\theta].}
\end{array}
\end{equation*}
Then, by carrying out brute force computations and by using  cohomological techniques, we obtained ${\cal B}[A^{(a,t\theta)}\!,t\theta\big]$ such that
\begin{equation*}
 t\>\frac{d}{dt}\> {\cal A}[A(a,t\theta),\Lambda(\lambda,t\theta),t\theta]
      = s_{NC}\, {\cal B}\,
            \big[A^{(a,t\theta)}\!,t\theta\big],
\end{equation*}
and, hence,
\begin{equation*}
  {\cal A}[A(a,\theta),\Lambda(\lambda,\theta),\theta] =  {\cal A}^{\rm Bardeen}  -
    s \int_0^1  \frac{dt}{t} ~
          {\cal B}[A(a,t\theta),t\theta].
\end{equation*}
We thus concluded that the $\theta$-dependent contributions to ${\cal A}[A(a,\theta),\Lambda(\lambda,\theta),\theta]$ are cohomologically trivial: they are not
anomalous contributions!

Since the previous result partially relies on the use a dimensionally
regularised Feynman integrals involving $\gamma_5$, it would be advisable to
check whether that result still holds for other regularization
methods. Another way  to obtain the gauge anomaly is Fujikawa's method: the
gauge anomaly signals that the fermionic measure is not invariant under chiral
gauge transformations. Fujikawa's method helps establish a connection with index theorems.
As yet, we lack a derivation of the absence of $\theta$-dependent anomalous terms by using Fujikawa's method.

Within Fujikawa's formalism, the ordinary gauge anomaly comes in two guises,
related by  local redefinitions of the corresponding currents: the consistent
form, ${\cal A}_{con}$, and the covariant form, ${\cal A}_{cov}$:
\begin{itemize}
   \item ${\cal A}_{con}$ verifies the Wess-Zumino consistency conditions and involves lengthy and tedious algebra. It is not gauge covariant.
   \item ${\cal A}_{cov}$ does not verify the Wess-Zumino conditions, it is gauge
     covariant and, as a result, the algebraic computations that
lead to it are simpler than in the "consistent" case.
\end{itemize}
As I was preparing a preliminary version of this talk, I decided to work out
the covariant form, up to first order in $\theta$,  of the gauge anomaly in
the U(1) case. Let me point out that the absence of $\theta$-dependent contributions to
the $U(1)$ gauge anomaly is nontrivial from the cohomological point of view of
Barnich, Brandt and Henneaux.  The results that I obtained are displayed
next.

Let me begin with the following partition function
\begin{equation*}
\begin{array}{l}
{Z[a,\theta] \;\equiv\;\int  d\bar{\psi}d\,\psi\quad e^{-\idx\bar{\psi}i{\cal D}\psi} }\\
{{\cal D}=\hat{\Dirac}+ \hat{\Rs},\quad \hatDirac=\prslash-i\aslash P_{L}}\\
{\hat{\Rs}=-[\frac{1}{4}\theta^{\alpha\beta}\,f_{\alpha\beta}
\gamma^\mu D_\mu +\frac{1}{2}\,\theta^{\alpha\beta}\,\gamma^\rho f_{\rho\alpha}
D_\beta]P_{L} }\\
\end{array}
\end{equation*}
Then, following Fujikawa, one introduces two bases of orthonormal  eigenfunctions $\{\varphi_m\}$ and $\{\phi_m\}$,
\begin{equation*}
\begin{array}{l}
{\Big(i{\cal D}(a)\Big)^{\dagger}i{\cal D}(a)\varphi_m\,=\,\lambda^2_m\,
\varphi_m,\quad i{\cal D}(a) \Big({i\cal D}(a)\Big)^{\dagger}\phi_m\,=\,
\lambda^2_m\,\phi_m,}
\end{array}
\end{equation*}
and expands
\begin{equation*}
\psi=\sum_{m}\,a_m\varphi_m,\quad\bar{\psi}=\sum_{m}\,\bar{b}_m{\phi}^{\dagger}_{\m},
\end{equation*}
which leads to the following definition of the fermionic measure:
\begin{equation*}
d\bar{\psi}d\,\psi\equiv \prod_{m}\,d\bar{b}_m da_m.
\end{equation*}

It is nor difficult to show that the gauge anomaly equation in covariant disguise reads
\begin{equation*}
\idx\text{Tr}\,\omega(x)
\big(D^{\mu}[a]{\cal J}_\mu^{(cov)}\big)(x)
=-\delta J \equiv {\cal A}[\omega,a,\theta]_{cov},
\end{equation*}
where
\begin{equation*}
\begin{array}{l}
{\delta J=d\bar{\psi}^{'}d\,\psi^{'}-d\bar{\psi}d\,\psi\,\quad
\psi^{'}=\psi+i\omega P_{L}\psi,\bar{\psi}^{'}=\bar{\psi}-i\bar{\psi}P_{R}\omega}\\[4pt]
{\delta J = \text{lim}_{\Lambda\rightarrow\infty}\idx \sum_m\{\phi^{\dagger}_m \omega e^{-\lambda^2_m/\Lambda^2} P_{R}\phi_m-
\varphi^{\dagger}_m\omega e^{-\lambda^2_m/\Lambda^2} P_{L}\varphi_m\}}\\[4pt]
{{\cal J}_\mu^{a,\,(cov)}(x)=\frac{1}{Z[a,\theta]}\int  d\bar{\psi}d\,\psi\, \frac{\delta S_{fermionic}}{\delta a^a_{\mu}(x)}\, e^{-S_{fermionic}},\,S_{fermionic}=\idx\bar{\psi}i{\cal D}\psi}.
\end{array}
\end{equation*}
By changing to a plane wave basis, one gets
\begin{equation*}
\begin{array}{l}
{{\cal A}[\omega,a,\theta]_{cov}=\mbox{lim}_{\Lambda\rightarrow\infty}\;
-\idx{\rm Tr}\;\omega(x)\idp\,{\rm tr}\;\Big\{
\Big(\gamma_{5}\,e^{-ipx}\,e^{-\frac{(i\Dirac^{(\theta)})(a))^2}{\Lambda^2}}\,e^{ipx}\Big)
\Big\},}\\[4pt]
{\Dirac^{(\theta)}(a)=\Dirac+\Rs,\Rs=-[\frac{1}{4}\theta^{\alpha\beta}\,f_{\alpha\beta}
\gamma^\mu D_\mu +\frac{1}{2}\,\theta^{\alpha\beta}\,\gamma^\rho f_{\rho\alpha}
D_\beta].}
\end{array}
\end{equation*}
Let us expand next the previous result in powers of $\theta$ and remove the terms that vanish as
$\Lambda\rightarrow\infty$. Thus one gets
\begin{equation*}
\begin{array}{l}
{  {\cal A}[\omega,a,\theta]_{cov}=\text{lim}_{\Lambda\rightarrow\infty}\;
-\idx{\rm Tr}\;\omega\idp\,{\rm tr}\;\Big\{
\Big(\gamma_{5}\,e^{-ipx}\,e^{-\frac{(i\Dirac^{(\theta)}(a))^2}{\Lambda^2}}\,e^{ipx}\Big)
\Big\}= }\\[4pt]
{ {\cal A}^{(ordinary)}[\omega,a]
+{\cal A}^{(1)}[\omega,a,\theta]+o(\theta^2) }\\[9pt]
{  {\cal A}[\omega,a]^{(ordinary)}=-\frac{1}{32\pi^2}\idx{\rm Tr}\omega\epsilon^{\mu\nu\rho\sigma}f_{\mu\nu} f_{\rho\sigma},}\\[4pt]
{{\cal A}^{(1)}[\omega,a,\theta]=\idx{\rm Tr}\omega(x)[{\cal A}_1(x)+{\cal A}_2(x)+{\cal A}_3(x)] }
\end{array}
\end{equation*}
\begin{equation*}
        \begin{array}{l}
                {{\cal A}_1=-\sum_{l=0}^{1}\limit 2i\idq e^{-q^2}\frac{1}{2}\text{tr}\gamma_5
      \Dp^{2l}(\Lambda q)\left\{\Dirac(\Lambda q),
  \Rs(\Lambda q)\right\}\Dp^{2(1-l)}(\Lambda q)\id,}\\[4pt]
     %%%%%
       { {\cal A}_2=-\sum_{l=0}^{2}\limit 2i\idq e^{-q^2}\frac{1}{3!\Lambda^2}\text{tr}\gamma_5
       \Dp^{2l}(\Lambda q)\left\{\Dirac(\Lambda q),\Rs(\Lambda q)\right\}\Dp^{2(2-l)}(\Lambda q)\id,}\\[4pt]
     %%%%%%
        {{\cal A}_3=-\sum_{l=0}^{3}\limit 2i\idq e^{-q^2}\frac{1}{4!\Lambda^4}\text{tr}\g_5
       \Dp^{2l}(\Lambda q)\left\{\Dirac(\Lambda q),\Rs(\Lambda q)\right\}\Dp^{2(3-l)}(\Lambda q)\id.}
    \end{array}
    \end{equation*}
Some lengthy algebra and the fact that the $a_\m$'s commute --U(1) case--
lead to
\begin{equation*}
\begin{array}{l}
   { {\cal A}_1=-\frac{i}{8\pi^2}\theta^{\alpha\beta}\epsilon^{\mu\nu\rho\sigma}
(-\frac{1}{2} f_{\alpha\beta}f_{\mu\nu}f_{\rho\sigma}-f_{\nu\alpha}
f_{\mu\beta}f_{\rho\sigma})}\\[4pt]
   {\phantom{{\cal A}_1}+\frac{1}{16\pi^2}\theta^{\alpha\beta}
\epsilon^{\mu\nu\rho\sigma}\big[f_{\mu\nu}(\partial_\rho f_{\sigma\alpha}
D_\beta \id+\frac{1}{2} \partial_\rho f_{\alpha\beta} D_{\sigma} \id)+\partial_\mu f_{\nu\alpha}
f_{\rho\sigma}D_\beta \id}\\[4pt]
   {\phantom{T_1=+\frac{1}{16\pi^2}\theta^{\alpha\beta}
\epsilon^{\mu\nu\rho\sigma}\big[}+\frac{1}{2} \partial_\mu f_{\alpha\beta} f_{\rho\sigma} D_\nu \id\big],}\\[4pt]
{ {\cal A}_2=-\frac{i}{2(4\pi)^2}\theta^{\alpha\beta}\epsilon^{\mu\nu\rho\sigma}
f_{\alpha\beta}f_{\mu\nu}f_{\rho\sigma}+}\\[4pt]
{\phantom{{\cal A}_2=}
-\frac{1}{16\pi^2}\theta^{\alpha\beta}\epsilon^{\mu\nu\rho\sigma}
[f_{\mu\nu}\partial_\rho f_{\sigma\alpha}D_\beta \id+\partial_\mu
  f_{\nu\alpha}f_{\rho\sigma}D_\beta \id+\frac{1}{2}
(f_{\mu\nu}\partial_\rho f_{\alpha\beta}D_\sigma\id}\\[4pt]
{\phantom{T_2^{(4\gamma)}=-\frac{1}{16\pi^2}\theta^{\alpha\beta}
\epsilon^{\mu\nu\rho\sigma}
[}+\partial_\mu f_{\alpha\beta}f_{\rho\sigma}D_\nu\id)],}\\[4pt]
{{\cal A}_3=0.}
\end{array}
\end{equation*}

So, finally the first order in $\theta$ correction to the ordinary anomaly vanishes:
\begin{equation*}
\begin{array}{l}
{{\cal A}^{(1)}[\omega,a,\theta]=\idx{\rm Tr}\omega(x)[{\cal A}_1(x)+{\cal A}_2(x)+
{\cal A}_3(x)]=}\\[4pt]
{\frac{i}{32\pi^2}\theta^{\alpha\beta}\epsilon^{\mu\nu\rho\sigma}\idx
\text{Tr}\omega(f_{\alpha\beta}f_{\mu\nu}f_{\rho\sigma}+4\,f_{\nu\alpha}
f_{\mu\beta}f_{\rho\sigma})=0,}\\[4pt]
{{\cal A}[\omega,a,\theta]_{cov}={\cal A}^{(ordinary)}[\omega,a]+o(\theta^2)}.
\end{array}
\end{equation*}
This shows complete agreement with the result obtained by using Dimensional
Regularization. Higher order corrections in $\theta$ and the nonabelian case
are still to be worked out.

\section{Renormalisability}

The issue of the renormalisability of noncommutative gauge theories formulated
within the enveloping-algebra formalism started off splendidly, for it was shown
by Bichl, Grimstrup, Grosse, Popp. Schweda and Wulkenhaar~\cite{Bichl:2001cq}
that the photon two-point function is renormalisable at any order in
$\theta$. Unfortunately, Wulkenhaar showed~\cite{Wulkenhaar:2001sq} that this
$\theta$-expanded  QED was not renormalisable mainly due to the infamous
four-point fermionic divergence:

\begin{equation*}
\frac{c}{D-4}\theta^{\alpha\beta}\epsilon_{\mu\nu\rho\sigma}
\idx\bar{\psi}\gamma_5\gamma^{\rho}\psi\bar{\psi}\gamma^{\sigma}\psi.
\end{equation*}

Four  years after Wulkenhaar's paper, there came along the encouraging results
by Buric, Latas and Radovanovic~\cite{Buric:2005xe}, and, Buric, Radovanovic and Trampetic~\cite{Buric:2006wm}, that the gauge sector
of SU(N) and the noncommutative Standard Model were one-loop renormalisable at
first order in $\theta$.  And yet, due to the infamous four-point fermionic
divergence above, the construction of theories with a renormalisable one-loop
and first-order-in-$\theta$ matter sector remained an open issue. Then it
appeared a new  paper by  Buric, Latas, Radovanovic and
Trampetic~\cite{Buric:2007ix}, where they showed that the divergence of the  four-point
fermionic function vanishes for a noncommutative SU(2) chiral theory with the
matter sector  being an SU(2)-doublet of noncommutative left-handed fermions.
This result was later generalized in ref.~\cite{Martin:2009sg} to any
noncommutative GUT-inspired theory with only fermions as matter fields. Let me
recall that by noncommutative GUT-inspired theories I mean gauge theories
whose noncommutative fermions are all --this is capital-- left-handed multiplets,
which transforms under a Grand Unification group. Thus, one of the
obstacles --what about the renormalisability of the other 1PI functions?-- to
achieve one-loop and first-order-in-$\theta$ renormalisability had been
removed by selecting Grand Unification --and, as we shall see, family unification, besides-- as a guiding principle.

The absence of the infamous four-point fermionic divergence opened up the possibility
of building noncommutative theories with massless fermionic noncommutative
chiral matter that are one-loop renormalisable at first order in $\theta$.
Actually, Wulkenhaar had already  pointed out in ref.~\cite{Wulkenhaar:2001sq}
that, in the massless case,  noncommutative QED is (off-shell) one-loop renormmalisable at first order in $\theta$,
if one forgets about the fermionic four-point function. At long last, it was
shown in ref.~\cite{Martin:2009vg} that noncommutative GUT-inspired theories,
with a matter sector made out of fermions and no scalars, were, on-shell and
at first order in $\theta$, one-loop-renormalisable for any anomaly safe
compact simple gauge group, if, and only if, all the flavour fermionic multiplets carry irreps with the same
quadratic Casimir, ie, renormalisability is very partial to  family
unification. This selects SO(10), $\text{E}_6$, and drops SU(5), as noncommutative
Grand Unification groups--see~\cite{Tamarit:2009iy}.

We shall close this section with a quick recap of the results in
ref.~\cite{Martin:2009vg}. The action of the noncommutative GUT-inspired
models in ref.~\cite{Martin:2009vg} reads
\begin{align*}
&{S}=\idx-\frac{1}{2g^2}\Tr F_{\m\n}\star F^{\m\n}+\Psib_L i\Dirac \Psi_L,\\[4pt]
%%%
&\nonumber F_{\m\n}=\partial_\m A_\n-\partial_\n A_\m-i[A_\m, A_\n]_\star,\quad D_{\m}\psi_L=\partial_\m\Psi_L-i
\rho_{\Psi}(A_\m)\star\Psi_L,
\end{align*}
$\rho_{\psi}$  denotes an arbitrary unitary representation, which is a direct sum of irreducible representations, $\rho_{\psi}=\bigoplus_{r=1}^F \rho^{r}_{\psi}$.
Then, lengthy computations led to the following result:

Once $\psi^r_L$, $g$ and $\theta$ have been renormalised as follows
\begin{equation*}
\begin{array}{l}
{\psi^r= (Z^r_\psi)^{1/2}\psi^r_{R}, g=\mu^{-\epsilon}Z_g g_R,  \theta^{\mu\nu}=Z_\theta \theta_R^{\mu\nu},}\\[4pt]
{Z^r_\psi=1+\frac{g^2 C_2(r)}{16\pi^2\epsilon},
 Z_g=1+\frac{g^2}{16\pi^2\epsilon}\Big[\frac{11}{6}C_2(G)-\frac{2}{3}\sum_r c_2(r)\Big],}\\[4pt]
 {Z_\theta=-Z^r_\psi-\frac{g^2}{48\pi^2\epsilon}(13C_2(r)-4C_2(G)),}
\end{array}
\end{equation*}
the UV divergences, at one-loop and first order in $\theta$, which remain in
the background-field effective action are given by the on-shell vanishing expression
\begin{equation*}
S^{\rm ct}=\idx \frac{\delta S}{\delta a_\m^a(x)} F_\m^a[a,\psi]+\Big(\sum_r\frac{\delta S}{\delta\psi^{r}(x)}G_r[a,\psi]+{\rm c.c.}\Big),
\end{equation*}
where
\begin{equation*}
\begin{array}{l}
{F_\mu=y_1\theta^{\alpha\beta}{\cal D}_\m f_{\alpha\beta}+y_2{\theta_\mu}^\alpha  D^\nu f_{\nu\alpha}+\sum_r y^r_3{\theta_\mu}^\alpha(\bar\psi_r\gamma_\alpha P_L T^a\psi^r)T^a}\\[4pt]
 { \phantom{F_\mu=}+i\sum_r y^r_4{\theta}^{\alpha\beta}(\bar\psi_r\gamma_{\mu\alpha\beta} P_L T^a\psi^r)T^a+y_5{\tilde{\theta}_\mu}^{\,\,\,\,\beta} D^\nu f_{\nu\beta},}\\[4pt]
{G_{r,L}=k^r_1\theta^{\alpha\beta}f_{\alpha\beta}P_L\psi^r+
k^2_r\theta^{\alpha\beta}{\gamma_{\alpha\mu}}P_L{f_\beta}^\mu \psi^r}\\[4pt]
{\phantom{G_{r,L}=}+k^r_3\theta^{\alpha\beta}\gamma_{\alpha\mu}P_L D_\beta D^\mu\psi^r+k^r_4\th^{\alpha\beta}\gamma_{\alpha\beta}P_L D^2\psi^r}\\[4pt]
{\phantom{G_{r,L}=}+k^r_5\tilde{\theta}^{\alpha\beta}\gamma_5 P_L f_{\alpha\beta}\psi^r; \,\,\,y_i\in\mathbb{R},\,k_i\in\mathbb{C},}
\end{array}
\end{equation*}
with
\begin{equation*}
\begin{array}{l}
 {y_1={\rm Im}k^r_1, y^r_3=2g^2 y_2,}\\[4pt]
 {y^r_4=-y_5 g^2-\frac{g^4}{384\pi^2}(16 C_2(r)-13 C_2(G)),} \\[4pt]
 {{\rm Re}k^r_1=-\frac{1}{2}
 {\rm  Im}k^r_3-\frac{g^2}{384\pi^2\epsilon}(13C_2(r)-8C_2(G)),}\\[4pt]
 {{\rm Im}k^r_5=-\frac{g^2}{384\pi^2\epsilon}(11C_2(r)-8C_2(G)),}\\[4pt]
 {{\rm Im}k^r_4 =\frac{g^2 C_2(r)}{384\pi^2\epsilon},
 {\rm  Re}k^r_2=-\frac{5g^2}{192\pi^2\epsilon}(2C_2(r)-C_2(G)),}\\[4pt]
 {{\rm Im}k^r_2={\rm Re}k^r_3=2{\rm Re}k^r_5=-2{\rm Re}k^r_4.}
 \end{array}
 \end{equation*}
Let me  stress that  $y_1,y_2,y_5$ and $Z_\theta$ above must be flavour
independent, and so must be $y_3,y_4$. Hence. $C_{2}(r)$, must be the same
  for all irreps, which in turn demands family unification.

\section{Yukawa Terms in Noncommutative GUTs}

Here I shall address the issue of constructing Yukawa terms in noncommutative
SO(10) and $\mbox{E}_6$ GUTs. For details I refer the reader to
ref.~\cite{Martin:2010ng}.

Let us recall that Yukawa terms of ordinary SO(10) and $\mbox{E}_6$ read
\begin{equation}
\calY^{\text{(ord)}}=\idx\;\funY_{ff'}\;\msC_{AiB}\;\tpsi^{\alpha}_{A  f}\;\psi_{\alpha B f'}
\;\phi_{i},
\label{ordyuk}
\end{equation}
where
$\tpsi^{\alpha}_{ f}\equiv (\psi^{\alpha}_{  f})^{t}$, and $\funY_{ff'}$
denotes the Yukawa coefficients. For SO(10), each fermionic multiplet
$\psi_{\alpha  f'}$ carry the 16 irrep of SO(10), whereas, in the
$\mbox{E}_6$ GUT, $\psi_{\alpha  f'}$ transforms under the 27 irrep of
$\mbox{E}_6$. The Higgs multiplets in SO(10) carry any of the
following irrreps: 10, 120 and $\overline{126}$. In the  $\mbox{E}_6$ case the
Higgs multiplets furnish any of the irreps of  $\mbox{E}_6$ that I enumerate
now: 27, 351' and 351. Indeed, one has the following Clebsch-Gordan decompositions
\begin{equation*}
 16\bigotimes 16=(10\bigoplus 126)_{\text{s}}\bigoplus 120_{\text{as}},
 27\bigotimes
27=(\overline{27}\bigoplus \overline{351'})_{\text{s}}
\bigoplus \overline{351}_{\text{as}}
 \end{equation*}

In eq.~(\ref{ordyuk}), $\msC_{AiB}$ is an invariant tensor:
\begin{equation*}
\tSigma^a_{AC}\,\msC_{CiB}\,+\,\msC_{AjB}
M^a_{ji}\,+\,\msC_{AjC}\,\Sigma^a_{CB}=0,
\end{equation*}
where  $\tSigma^a$, $M^a$ and $\Sigma^a$  denote the group generators in the
irreps furnished by $\tpsi^{\alpha}_{A f}$,
$\phi_i$ and $ \psi^{\alpha}_{B  f'}$, respectively.

Let $\tPsi^{\alpha}_{A f}$, $ \Psi_{\alpha B f'}$ and $\Phi_{i}$
denote the noncommutative fermionic and Higgs fields defined by standard Seiberg-Witten
maps, ie, solutions to
\begin{equation*}
 s_{NC}(NCField)\equiv i\Lambda\star(NCField)=s(NCField)
\end{equation*}
that match the ordinary fields at $\theta=0$. Then, a naive noncommutative version
\begin{equation*}
\calY^{(NC)}_{(naive)}=\idx\;\funY_{ff'}\;\msC_{AiB}\;\tPsi^{\alpha}_{A
  f}\;\star\;\Psi_{\alpha B f'}\;\star\;
\Phi_{i}
\end{equation*}
of the ordinary Yukawa term in eq.~(\ref{ordyuk}) would not do! Indeed,
\begin{equation*}
\begin{array}{l}
{0\neq s_{NC}\calY^{(NC)}_{(naive)}=}\\[2pt]
{\idx\;\funY_{ff'}\;\msC_{AiB}\;(i\tLambda_{AC}\star\tPsi^{\alpha}_{C
  f}\;\star\;\Psi_{\alpha B f'}\;\star\;
\Phi_{i}
+\tPsi^{\alpha}_{A
  f}\;\star\;i\Lambda_{BC}\star\Psi_{\alpha C f'}\;\star\;
\Phi_{i}}\\[2pt]
{\phantom{0\neq s_{NC}\calY^{(NC)}_{(naive)}=
\idx\;\funY_{ff'}\;\msC_{AiB}\;(}
+\tPsi^{\alpha}_{A
  f}\;\star\;\Psi_{\alpha B f'}\;\star\;i\Lambda_{ij}\star
\Phi_{j}),}
\end{array}
\end{equation*}
for the $\star$-product is not commutative and
$\msC_{AiB}$ is not invariant for enveloping-algebra valued $\Lambda$'s.

I shall now explain my  strategy for constructing  noncommutative Yukawa terms.
To carry over the properties of $\msC_{AiB}$ to the noncommutative theory in a
consistent way, one first combines $\msC_{AiB}$ with the ordinary fields
$\tpsi^{\alpha}_{A  f}$, $\psi_{\alpha B f'}$ and $\phi_{i}$, and, then,
defines new ordinary fields that transform under tensor products of ordinary
irreps of the gauge group, but have the  very same number of freedom as $\tpsi^{\alpha}_{A  f}$, $\psi_{\alpha B f'}$ and $\phi_{i}$:
\begin{equation*}
\phi_{AB}=\msC_{AiB}\,\phi_{i},\quad\tpsi^{\alpha}_{iBf}=\tpsi^{\alpha}_{Af}\;
\msC_{AiB},\quad \psi_{\alpha Aif'}=\msC_{AiB}\;\psi_{\alpha Bif'}.
\end{equation*}
The BRS transformations of these new fields run thus:
\begin{equation*}
\begin{array}{l}
{s\phi_{AB}=-i\,\tlambda^{(\psi)}_{AC}\,\phi_{CB}\,-\,i\,\phi_{AC}\,
\lambda^{(\psi)}_{CB},}\\[4pt]
{s\tpsi^{\alpha}_{iBf}=-i\,\tlambda^{(\phi)}_{ij}\,\tpsi^{\alpha}_{jBf}\,-\,i\,
\tpsi^{\alpha}_{iCf}\,\lambda^{(\psi)}_{CB},}\\[4pt]
{s\psi_{\alpha Aif'}=-i\,\tlambda^{(\psi)}_{AC}\,\psi_{\alpha Cif'}\,-\,i\,
\psi_{\alpha Ajf'}\lambda^{(\phi)}_{ji}.}\\
\end{array}
\end{equation*}

Next, to each ordinary field $\phi_{AB}$, $\tpsi^{\alpha}_{iBf}$ and
$\psi_{\alpha Aif'}$, one associates a noncommutative counterpart
\begin{equation*}
\Phi_{AB}[\phi_{AB},a^a_{\mu},\theta],\quad \tPsi^{\alpha}_{iBf}[\tpsi^{\alpha}_{iBf},a^a_{\mu},\theta]\quad\text{and}\quad
\Psi_{\alpha Aif'}[\psi_{\alpha Aif'},a^a_{\mu},\theta],
\end{equation*}
 which, respectively, are solutions to the following Hybrid Seiberg-Witten map equations:
\begin{equation}
s_{NC}\Phi_{AB}=s\Phi_{AB},\quad
s_{NC}\tPsi^{\alpha}_{iBf}=s\tPsi^{\alpha}_{iBf},\quad
s_{NC}\Psi_{\alpha Aif'}=s\Psi_{\alpha Aif'},
\label{hysweq}
\end{equation}
where one defines
\begin{equation*}
\begin{array}{l}
{s_{NC}\Phi_{AB}\equiv -i\,
\tLambda^{(\psi)}_{AC}
\star\Phi_{CB}-i\,\Phi_{AC}\star
\Lambda^{(\psi)}_{CB}},\\[2pt]
{s_{NC}\tPsi^{\alpha}_{iBf}\equiv-i\,\tLambda^{(\phi)}_{ij}\star \tPsi^{\alpha}_{jBf}-i\,\tPsi^{\alpha}_{iCf}\star
\Lambda^{(\psi)}_{CB}}\\[2pt]
{s_{\text{nc}}\Psi_{\alpha Aif'}\equiv-i\,\tLambda^{(\psi)}_{AC}\star \Psi_{\alpha Cif'}-i\,\Psi_{\alpha Ajf'}\star\Lambda^{(\phi)}_{ji}}.
\end{array}
\end{equation*}
Let me point out that the  action from the left and from the right (as opposed to both actions from
the left or both from the right) of the $\Lambda$'s is the only choice consistent
with  $(s_{NC})^2=0$!.
The solutions to the equations in eq.~(\ref{hysweq}) are Seiberg-Witten maps of hybrid type, a notion introduced
by Schupp~\cite{Schupp:2001we}.

We are now in the position to obtain in a natural (naive) way noncommutative
SO(10), $\text{E}_6$ Yukawa terms from their ordinary counterparts. Indeed, in terms of $\phi_{AB}$, the ordinary Yukawa term reads:
\begin{equation*}
 \calY_1^{\text{(ord)}}\equiv \calY^{\text{(ord)}}=
\idx\;\funY_{ff'}\;\tpsi^{\alpha}_{ A  f}\;\phi_{AB}\;\psi_{\alpha B f'},
\end{equation*}
so that, its noncommutative counterpart is
\begin{equation*}
\calY_1^{\text{(nc)}}=
\idx\;\funY^{(1)}_{ff'}\;\tPsi^{\alpha}_{ A  f}\star\Phi_{AB}\star\Psi_{\alpha B f'}.
\end{equation*}
In words: the noncommutative Yukawa term associated to
$\calY_1^{\text{(ord)}}$ is obtained by replacing each ordinary field in the
latter with its noncommutative counterpart and the ordinary product with the $\star$-product.

By construction $\calY_1^{\text{(nc)}}$ is invariant under the following noncommutative BRS transformations:
\begin{equation*}
 \begin{array}{l}
 {s_{NC}\tPsi^{\alpha}_{ A  f}=i\,\tPsi^{\alpha}_{B f}\star \tLambda^{(\psi)}_{BA},\quad s_{\text{nc}}\Psi_{\alpha B f'}=i\,\Lambda^{(\psi)}_{BC}\star\Psi_{\alpha C f'},}\\[4pt]
 {s_{NC}\Phi_{AB}=-i\,\tLambda^{(\psi)}_{AC}\star\Phi_{CB}-i\,\Phi_{AC}\star
\Lambda^{(\psi)}_{CB},}\\[4pt]
{s_{NC}\tLambda^{(\psi)}_{BA}=-i\,\tLambda^{(\psi)}_{BC}\star\tLambda^{(\psi)}_{CA},
\quad s_{\text{nc}}\Lambda^{(\psi)}_{BC}=i\,\Lambda^{(\psi )}_{BD}\star\Lambda^{(\psi)}_{DC}.}\\
\end{array}
\end{equation*}
The Seiberg-Witten maps which define the noncommutative fields are
\begin{equation*}
\begin{array}{l}
{\tPsi^{\alpha}_{ A f}=\tpsi^{\alpha}_{ A f}-\frac{1}{2}\,\theta^{\mu\nu}\,
\partial_{\mu}\tpsi^{\alpha}_{ B f}\widetilde{a}^{(\psi)}_{\nu\,BA}+
\frac{i}{4}\,\theta^{\mu\nu}\,\tpsi^{\alpha}_{ C f}\widetilde{a}^{(\psi)}_{\mu\,CB}
\widetilde{a}^{(\psi)}_{\nu\,BA}+O(\theta^2),}\\[4pt]
{\Phi_{ A B}=\phi_{ A B}+\frac{1}{2}\,\theta^{\mu\nu}\,
\widetilde{a}^{(\psi)}_{\mu\,AC}\partial_{\nu}\phi_{CB}+
\frac{i}{4}\,\theta^{\mu\nu}\,\widetilde{a}^{(\psi)}_{\mu\,AC}
\widetilde{a}^{(\psi)}_{\nu\,CD}\phi_{DB}+}\\[4pt]
{\phantom{\Phi_{AB}=\phi_{AB}}+\frac{1}{2}\,\theta^{\mu\nu}\,
\partial_{\mu}\phi_{AC}a^{(\psi)}_{\nu\,CB}+
\frac{i}{4}\,\theta^{\mu\nu}\,\phi_{AC}a^{(\psi)}_{\mu\,CD}
a^{(\psi)}_{\nu\,DB}}\\[4pt]
{\phantom{\Phi_{AB}=\phi_{AB}}+\frac{i}{2}\,\theta^{\mu\nu}\,
\widetilde{a}^{(\psi)}_{\mu\,AC}\phi_{CD}a^{(\psi)}_{\nu\,DB}+O(\theta^2),}\\[4pt]
{\Psi_{\alpha B f'}=\psi_{\alpha B f'}-\frac{1}{2}\,\theta^{\mu\nu}\,
a^{(\psi)}_{\mu\,BC}\partial_{\mu}\psi_{\alpha C f'}+
\frac{i}{4}\,\theta^{\mu\nu}\, a^{(\psi)}_{\mu\,BC}a^{(\psi)}_{\nu\,CD}\psi^{\alpha}_{ D f'}+O(\theta^2).}
\end{array}
\end{equation*}

Let me now point out that if we use $\tpsi^{\alpha}_{iBf}$ and
$\psi_{\alpha Aif'}$ to formulate an ordinary Yukawa term, we obtain
the same ordinary Yukawa term:
\begin{equation*}
\begin{array}{l}
{\calY_2^{\text{(ord)}}=
\idx\;\funY_{ff'}\;\tphi_i\;\tpsi^{\alpha}_{i B  f}\;\psi_{\alpha B f'},}\\[4pt]
{\calY_3^{\text{(ord)}}=
\idx\;\funY_{ff'}\;\tpsi^{\alpha}_{A  f}\;\psi_{\alpha A i f'}\;\phi_i},\\[4pt]
{\calY_1^{\text{(ord)}}=\calY_2^{\text{(ord)}}=\calY_3^{\text{(ord)}}.}
\end{array}
\end{equation*}
And yet, the noncommutative counterparts of $\calY_2^{\text{(ord)}}$ and
$\calY_3^{\text{(ord)}}$ are not equal:
\begin{equation*}
\begin{array}{l}
{\calY_2^{\text{(nc)}}=
\idx\;\funY^{(2)}_{ff'}\;\tPhi_{i}\star\tPsi^{\alpha}_{ iB  f}\star\Psi_{\alpha B f'}}\\[4pt]
{\calY_3^{\text{(nc)}}=
\idx\;\funY^{(3)}_{ff'}\;\msC_{AiB}\;\tpsi^{\alpha}_{ A  f}\;\phi_i\;\psi_{\alpha B f'}}\\[4pt]
{\calY_1^{\text{(nc)}}\neq\calY_2^{\text{(nc)}}
\neq\calY_3^{\text{(nc)}}\neq\calY_1^{\text{(nc)}}}
\end{array}
\end{equation*}
Hence, I propose the following  noncommutative Yukawa term
\begin{equation*}
\calY^{\text{(nc)}}=\calY_1^{\text{(nc)}}\,+\,\calY_2^{\text{(nc)}}\,+\,
\calY_3^{\text{(nc)}}.
\end{equation*}
It can be shown --see ref~\cite{Martin:2010ng}-- that at first order in
$\theta$ this is the most general BRS invariant Yukawa-type term 
\begin{equation*}
\theta^{\mu\nu}\idx\calY_{ff'}\;\psi^{\alpha}_{Af}\;{\cal V}^{AiB}_{\mu\nu}[\theta^{\rho\sigma},\partial_{\mu},a^a_{\nu}]\;\phi_i\;\psi_{\alpha Bf'}
\end{equation*}
that one can write. This Yukawa term is therefore renormalisable at first order in $\theta$.

\section{What about SUSY?}

For $U(N)$  in the fundamental rep., ${\cal N}=1$ SYM exists  in the enveloping-algebra formalism as a classical theory:
\begin{equation*}
S_{NCSYM}=\frac{1}{2g^2}\text{Tr}\idx [-\frac{1}{2}\,F^{\mu\nu}\star F_{\mu\nu}-2i\,\Lambda^{\a}\star\sigma^{\mu}_{\a\,\dot{\a}}D_{\m}\bar{\Lambda}^{\dot{\a}}+D\star D]
\end{equation*}
where
\vskip -0.4cm
\begin{equation*}
A_\mu=A_\mu[a,\lambda_\alpha,d,\theta],\, \Lambda_{\alpha}[a,\lambda_\alpha,d,\theta]\, \text{and}\,
D=D[a,\lambda_\alpha,d,\theta]
\end{equation*}
are SW maps. $S_{NCSYM}$ is invariant under ${\cal N}=1$ SUSY:
\begin{itemize}
\item linearly realized in terms of the noncommutative fields
                     ( there is a local superfield formulation)\\
                      and
\item nonlinearly realized in terms of the ordinary fields
       (no local superfield formulation exists, but a nonlocal one does, at least for U(1) --see~\cite{Martin:2008xa}).
\end{itemize}
The ${\cal N}=1$ SUSY transformations of the noncommutative fields read
\begin{equation*}
\begin{array}{l}
{A_\mu[\varphi,\theta]\rightarrow A_\mu^{(\epsilon)}[\varphi,\theta]=
A_\mu[\varphi,\theta]+\delta_{\epsilon}A_\mu[\varphi,\theta]}\\
{\Lambda_{\alpha}[\varphi,\theta]\rightarrow  \Lambda_{\alpha}^{(\epsilon)}[\varphi,\theta]=
\Lambda_{\alpha}[\varphi,\theta]+\delta_{\epsilon}\Lambda_{\alpha}[\varphi,\theta]}\\
{D[\varphi,\theta]\rightarrow D^{(\epsilon)}[\varphi,\theta]=
D[\varphi,\theta]+\delta_{\epsilon}D[\varphi,\theta]}\\
\end{array}
\end{equation*}
where $\varphi$ denotes generically the ordinary fields and
\begin{equation*}
\begin{array}{l}
\delta_{\epsilon}A^\m=i\epsilon^{\a}\sigma^\m_{\a\,\dot{\a}}\bar{\Lambda}^{\dot{\a}}+
i\bar{\epsilon}^{\dot{\a}}\bar{\sigma}^\m_{\dot{\a}\,\a}\Lambda^\a,\\
\delta_{\epsilon}\Lambda_{\a}={(\sigma^{\m\n})_{\a}}^{\beta}\epsilon_{\beta}F_{\m\n}+i\epsilon_{\a}D,\\
\delta_{\epsilon}D=-\epsilon^{\a}\sigma^{\mu}_{\a\,\dot{\a}}D_{\m}\bar{\Lambda}^{\dot{\a}}+
\bar{\epsilon}^{\dot{\a}}\bar{\sigma}^{\mu}_{\dot{\a}\,\a}D_{\m}\Lambda^{\a}.\\
\end{array}
\end{equation*}
Now, the SUSY transformations have just introduced --do not forget that we
are in the $U(N)$ case in the fundamental representation-- can be induced by
performing a nonlinear variation of the ordinary fields, which up to first order in $\theta$, reads
\begin{equation}
\begin{array}{l}
{\delta_{\eps}  a_\m= \frac{1}{4}\eps\sigma_\m\bar{\lambda}-\frac{1}{4}
\bar\eps\bar\sigma_\m\lambda+\frac{1}{16}\theta^{\n\rho}\;\Big[
\{ a_\n,2D_\rho (\eps\sigma_\m\bar{\lambda}\!-\!\bar\eps\bar\sigma_{\m}\lambda)\!
-\!i[ a_\rho,\eps\sigma_\m\bar{\lambda}\!-\!\bar\eps\bar\sigma_\m\lambda]\}}\\
{\phantom{\delta_{\eps}  a_\m=} -\{\eps\sigma_n\bar{{\lambda}}\!-\!\bar\eps\bar\sigma_\n\lambda,\partial_\rho a_\m+\f_{\rho\m}\}\!
 -\!\{ a_\n,\partial_\rho (\eps\sigma_\m\bar{{\lambda}}\!-\!\bar\eps\bar\sigma_\m\lambda)\!+
 \!D_\rho(\eps\sigma_\m\bar{{\lambda}}\!-\!\bar\eps\bar\sigma_{m}\lambda)\!
}\\
 {\phantom{\delta_{\eps}  a_\m=} -\!D_\m(\eps\sigma_\rho\bar{{\lambda}}\!-\!\bar\eps\bar\sigma_{l}\lambda)\}  \Big]\,+\,\theta^2},\\
{\delta_{\eps}\lambda_{\alpha}=-\epsilon_{\alpha} d+ 2i\eps_\ga{(\s^{\m\n})^\g}_\a f_{\m\n}+
\frac{1}{4}\theta^{\n\rho}\;\Big[-\!\frac{1}{4}\{\eps\sigma_\n\bar{{\lambda}}\!-
\!\bar\eps\bar\sigma_\n\lambda,2D_\rho\lambda_{\alpha}\!-
\!i[a_\rho,\lambda_{\alpha}]\}}\\
{\phantom{\delta_{\eps} \lambda_{\alpha}=}\!-\!\{ a_\n,4iD_\rho(\eps_\ga{(\s^{\m\rho})^\g}_\a f_{\m\la})
+\!2[ a_\rho,\eps_\ga{(\s^{\m\la})^\g}_\a  f_{\m\la}]\!+\!\frac{i}{4}
[\eps\sigma_\rho\bar{{\lambda}}\!-\!\bar\eps\bar\sigma_l\lambda,\lambda_\a]\}\Big]}\\
{\phantom{\delta_{\eps} \lambda_{\alpha}=}+\theta^2}\\
{\delta_{\eps}  d= i\epsb\sb^\m D_\m\lambda+i\eps\s^\m D_\m
\bar{\lambda}\;+\frac{1}{4}\theta^{\n\rho}\,\Big[2i\{f_{\m\n},\bar\eps\bar\s^\m D_\rho\la+\eps\s^\m D_\rho\bar{\la}\} }\\
{\phantom{\delta_{\eps} d=}+i\{a_\n,(\partial_\rho +D_\rho)(\bar\eps\bar\s^\m D_\m\la+\eps\s^\m D_\m\bar{\la})\}
\!-\!\frac{1}{4}\{\eps\sigma_\n\bar{\lambda}\!-\!\bar\eps\bar\sigma_\n\lambda,2D_\rho d\!-\!i[ a_\rho,d]\}}\\
{\phantom{\delta_{\eps} d=}\!-\!\{ a_\n,2D_\rho(i\epsb\sb^\m D_\m
\lambda+i\eps\s^\m D_\m
\bar{\lambda})\!-\!i[ a_\rho,i\epsb\sb^\m D_\m\lambda+i\eps\s^\m D_\m
\bar{\lambda}]}\\
{\phantom{\delta_{\eps} d=}
\!+\!\frac{i}{4}[\eps\sigma_\rho\bar{\lambda}\!-
\!\bar\eps\bar\sigma_\rho\lambda, d]\}\Big]\,+\,\theta^2.}\\
\end{array}
\label{nlinearsusy}
\end{equation}
The following comments concerning the nonlinear variations of the ordinary
fields in eq.~(\ref{nlinearsusy}) are now in order:
\begin{itemize}
\item They are truly ${\cal N}=1$ SUSY transformations,
\begin{equation*}
[\delta_{\eps_2},\delta_{\eps_1}](\text{fields})=
 i(\eps_2\s^\m\epsb_1-\eps_1\s^\m\epsb_2)\partial_\m (\text{fields})+
 \text{gauge transformations},
 \end{equation*}
 due to the fact that the noncommutative fields carry a linear realisation of
 ${\cal N}=1$ SUSY. This holds at any order in $\theta$
--see ref.~\cite{Martin:2008xa}.
 \item $\delta_{\eps}  a_\m$, $\delta_{\eps}  \la_\a$ and $\delta_{\eps}  d$
   belong to the Lie algebra of the ordinary gauge group only
for U(N) in the
     fundamental rep. and its siblings, i.e.,
 \item for an arbitrary Lie algebra they take values on the enveloping-algebra: they are not  ordinary field variations which are also  ordinary fields.

\end{itemize}

The question that one should ask next is whether we can have SUSY
noncommutative GUTs. It is apparent that for simple gauge groups in any representation, it still makes
sense to consider the theory defined by the action
\begin{equation*}
S=\frac{1}{2g^2}\text{Tr}\idx [-\frac{1}{2}\,F^{\mu\nu}\star F_{\mu\nu}-2i\,\Lambda^{\a}\star\sigma^{\mu}_{\a\,\dot{\a}}D_{\m}\bar{\Lambda}^{\dot{\a}}+D\star D]
\end{equation*}
where
\begin{equation*}
A_\mu=A_\mu[a,\lambda_\alpha,d,\theta],\, \Lambda_{\alpha}[a,\lambda_\alpha,d,\theta]\quad\text{and}\quad
D=D[a,\lambda_\alpha,d,\theta]
\end{equation*}
are Seiberg-Witten maps. This action looks like a SUSY invariant
noncommutative action, for it is invariant under the following transformations
\begin{equation}
\begin{array}{l}
{A_\mu[\varphi,\theta]\rightarrow A_\mu^{(\epsilon)}[\varphi,\theta]=
A_\mu[\varphi,\theta]+\delta_{\epsilon}A_\mu[\varphi,\theta,]}\\
{\Lambda_{\alpha}[\varphi,\theta]\rightarrow  \Lambda_{\alpha}^{(\epsilon)}[\varphi,\theta]=
\Lambda_{\alpha}[\varphi,\theta]+\delta_{\epsilon}\Lambda_{\alpha}[\varphi,\theta],}\\
{D[\varphi,\theta]\rightarrow D^{(\epsilon)}[\varphi,\theta]=
D[\varphi,\theta]+\delta_{\epsilon}D[\varphi,\theta],}\\
{\delta_{\epsilon}A^\m=i\epsilon^{\a}\sigma^\m_{\a\,\dot{\a}}\bar{\Lambda}^{\dot{\a}}+
i\bar{\epsilon}^{\dot{\a}}\bar{\sigma}^\m_{\dot{\a}\,\a}\Lambda^\a},\\
{\delta_{\epsilon}\Lambda_{\a}={(\sigma^{\m\n})_{\a}}^{\beta}\epsilon_{\beta}F_{\m\n}+i\epsilon_{\a}D},\\
{\delta_{\epsilon}D=-\epsilon^{\a}\sigma^{\mu}_{\a\,\dot{\a}}D_{\m}\bar{\Lambda}^{\dot{\a}}+
\bar{\epsilon}^{\dot{\a}}\bar{\sigma}^{\mu}_{\dot{\a}\,\a}D_{\m}\Lambda^{\a},}\\
\end{array}
\label{offsw}
\end{equation}
and these transformations satisfy the ${\cal N}=1$ SUSY algebra commutation
relationship
\begin{equation*}
[\delta_{\eps_2},\delta_{\eps_1}](\text{NCfields})=
 i(\eps_2\s^\m\epsb_1-\eps_1\s^\m\epsb_2)\partial_\m (\text{NCfields})+
 \text{NC gauge transformations}.
 \end{equation*}
Notice that $\varphi$ denotes the ordinary fields $a_\mu$,
$\lambda_{\alpha}$ and $d$, and NC stands for noncommutative. This all  goes in
the right direction, but there is a catch:   $A_\mu^{(\epsilon)}[\varphi,\theta]$,
$\Lambda^{(\epsilon)}_{\alpha}[\varphi,\theta]$ and $D^{(\epsilon)}[\varphi,\theta]$ are not Seiberg-Witten maps in the sense that
there are no ordinary fields $\varphi^{(\epsilon)}[\varphi,\partial,\theta]$,
\begin{equation*}
\varphi^{(\epsilon)}[\varphi,\partial,\theta]=\varphi+\epsilon^{\alpha}\phi_{\alpha}[\varphi,\partial,\theta]+
\bar{\epsilon}_{\dot{\alpha}}\bar{\varphi}^{\dot{\alpha}}[\varphi,\partial,\theta],
\end{equation*}
such that
\begin{equation*}
A_\mu^{(\epsilon)}[\varphi,\theta]=A_\mu[\varphi^{(\epsilon)}[\varphi,\partial,\theta];\theta],\,
\Lambda^{(\epsilon)}_{\alpha}[\varphi,\theta]=\Lambda_{\alpha}[\varphi^{(\epsilon)}[\varphi,\partial,\theta];\theta],\,
D^{(\epsilon)}[\varphi,\theta]=D[\varphi^{(\epsilon)}[\varphi,\partial,\theta];\theta],
\end{equation*}
where $A_\mu[\varphi;\theta]$, $\Lambda_{\alpha}[\varphi;\theta]$ and
$D[\varphi;\theta]$ are solutions to the Seiberg-Witten equations satisfying
$A_\mu[\varphi;\theta=0]=a_\mu$, $\Lambda_{\alpha}[\varphi;\theta=0]=\lambda_{\alpha}$ and
$D[\varphi;\theta=0]=d$. The transformations in eq.~(\ref{offsw}) are
therefore defined from the space of noncommutative "physical" fields --those
defined by the Seiberg-Witten map as explained above--- into the space of general
fields taking values on the enveloping algebra. The so remaining question is
whether this invariance has any physical consequences.
In this regard, it is worth noticing that --unlike in the U(N) case-- the SUSY
noncommutative SU(N) theory thus obtained is one-loop and
first-order-in-$\theta$ (off-shell) renormalisable. This would be just a lucky
chance unless there is a symmetry at work, at first order in $\theta$, that
relates the gluon and gluino dynamics --see~\cite{Martin:2009mu}.

Some additional information regarding noncommutative SUSY theories defined by
means of the Seiberg-Witten map can be found in refs.~\cite{Paban:2002dr, Putz:2002ib,
Dayi:2003ju} and ~\cite{Mikulovic:2003sq}.

\section{Open problems}

We shall conclude with a list of pressing problems:
\begin{itemize}
\item For SO(10) and ${\rm E}_6$, inclusion of a phenomenologically relevant noncommutative Higgs potential: a  non trivial issue as implied by the construction of Yukawa terms.

\item Study of the one-loop renormalisability of those noncommutative GUTS at first order in $\theta$.

\item Construction and analysis of the properties of noncommutative SUSY SO(10), ${\rm E}_6$.

\item Study of the phenomenological implications of noncommutative SO(10), ${\rm E}_6$ GUTs.

\item Gauge anomalies, Fujikawa's method and index theorems. Recall that the index theorem in 2n+2 dimensions gives the gauge anomaly in 2n dimensions, that the index of the Dirac operator does not change under small deformations of it, and that in our formalism we are considering small deformations of the ordinary Dirac operator. Putting it all together: no $\theta$-dependent anomalous terms.

\item A challeging question: Will these noncommutative GUTs eventually find accommodation within F-theory~\cite{Cecotti:2009zf}?

\item A final question: can one formulate noncommutative GUTs
  without using the enveloping-algebra formalism? In answering this question 
  in the affirmative, the ideas presented in refs.~\cite{Grosse:2010zq,
    Chatzistavrakidis:2010xi} look most promising; see also~\cite{Bonora:2000td} .

\end{itemize}

\section{Acknowledgements}

I should like to thank the organizers --D. Bahns, H. Grosse and G. Zoupanos---
of this Workshop for giving me the opportunity to present the material covered
in this talk at this wonderful conference held in a  place with such beautiful
surroundings. This work has been financially supported in part by MICINN
through Grant No. FPA2008-04906.

\end{document}